\documentclass[Svgc]{Svgc}
\usepackage{graphics,dsfont,times,wrapfig,subfigure,latexsym,amsmath}

\journalname{Graphs and Combinatorics}

\renewcommand{\epsilon}{\varepsilon}
\newcommand{\R}{\ensuremath{\mathds R}}

\newcommand{\rightharpoonupfill}{{$\mathsurround=0pt\mathord-\mkern-6mu%
  \cleaders\hbox{$\mkern-2mu\mathord-\mkern-2mu$}\hfill
  \mkern-6mu\mathord\rightharpoonup$}}
\newbox\myrhpmbox\setbox\myrhpmbox=\hbox{$\rightharpoonup$}
\newbox\myrhpobox\newbox\myrhptbox
\newcommand{\overrightharpoonup}[1]{\setbox\myrhpobox=\hbox{\ensuremath{#1}}\ifdim\wd\myrhpmbox>\wd\myrhpobox\setbox\myrhptbox=\hbox to \wd\myrhpmbox{\hfil\box\myrhpobox\hfil}\setbox\myrhpobox=\box\myrhptbox\setbox\myrhptbox=\copy\myrhpmbox\else\setbox\myrhptbox=\hbox to \wd\myrhpobox{\rightharpoonupfill}\fi\vbox{\baselineskip=0pt\box\myrhptbox\kern-0.75\ht\myrhpmbox\box\myrhpobox}}

\DeclareMathOperator{\ma}{op}
\DeclareMathOperator{\CH}{CH}

\let\doendproof\endproof
\renewcommand\endproof{~\hfill\qed\doendproof}

\spnewtheorem{obs}{Observation}{\bfseries}{\itshape}
\spnewtheorem{conj}{Conjecture}{\bfseries}{\itshape}
\spnewtheorem{myclaim}{Claim}{\bfseries}{\itshape}

\graphicspath{{figures/}}

\begin{document}

\title{Maximizing Maximal Angles\\ for Plane Straight-Line
  Graphs\thanks{
    O.A., T.H., and B.V. were supported by the Austrian
            FWF National Research Network 'Industrial Geometry' S9205-N12. 
    C.H. was partially supported by projects MEC~MTM2006-01267 and 
        Gen.~Cat.~2005SGR00692. 
    A.P. was partially supported by Hungarian National Foundation Grant~T60427. 
    F.S. was partially supported by grant~MTM2005-08618-C02-02 of the Spanish 
        Ministry of Education and Science.
    B.S. was supported by the Netherlands Organisation for Scientific Research 
        (NWO) under project no.~639.022.707. 
  An extended abstract of this article has been presented in~\cite{AHHHSSV}.}}
\author{Oswin Aichholzer\inst{1} \and Thomas Hackl\inst{1} \and
  Michael Hoffmann\inst{2} \and Clemens Huemer\inst{3} \and Attila
  P{\'o}r\inst{4} \and Francisco Santos\inst{5} \and Bettina Speckmann\inst{6} \and Birgit Vogtenhuber\inst{1}
}                     

\institute{Institute for Software Technology, Graz University of Technology,\\ \email{[oaich|thackl|bvogt]@ist.tugraz.at}
\and Institute for Theoretical Computer Science, ETH Z\"urich,\\ \email{hoffmann@inf.ethz.ch}
\and Departament de Matem\`atica Aplicada IV, Universitat Polit\`ecnica de Catalunya,\\ \email{clemens.huemer@upc.edu}
\and Dept. of Appl. Mathem. and Inst. for Theoretical Comp. Science,
Charles Univ.,\\ \email{por@kam.mff.cuni.cz}
\and Dept. de Matem\'{a}ticas, Estad\'{\i}stica y Computaci\'{o}n, Universidad de Cantabria,\\ \email{francisco.santos@unican.es}
\and Department of Mathematics and Computer Science, TU Eindhoven,\\ \email{speckman@win.tue.nl}}


\maketitle

\begin{abstract}
Let $G=(S, E)$ be a plane straight-line graph on a finite point set $S\subset\R^2$ in general position. The \emph{incident angles} of a point $p \in S$ in $G$ are the angles between any two edges of $G$ that appear consecutively in the circular order of the edges incident to $p$. A plane straight-line graph is called \textit{$\varphi$-open} if each vertex has an incident angle of size at least $\varphi$. In this paper we study the following type of question: What is the maximum angle $\varphi$ such that for any finite set $S\subset\R^2$ of points in general position we can find a graph from a certain class of graphs on $S$ that is $\varphi$-open? In particular, we consider the classes of triangulations, spanning trees, and spanning paths on $S$ and give tight bounds in most cases.
\end{abstract}


\section{Introduction}
Conditions on angles in plane straight-line graphs have been studied extensively in discrete and computational geometry. It is well known that Delaunay triangulations maximize the minimum angle over all triangulations, and that in a (Euclidean) minimum weight spanning tree each angle is at least $\frac{\pi}{3}$. In this paper we address the fundamental combinatorial question, what is the maximum value $\varphi$ such that for each finite point set in general position there exists a (certain type of) plane straight-line graph where each vertex has an incident angle of size at least $\varphi$. In other words, we consider $\min$-$\max$-$\min$-$\max$ problems, where we minimize over all finite point sets $S$ in general position in the plane, the maximum over all plane straight-line graphs $G$ (of the considered type), of the minimum over all $p \in S$, of the maximum angle incident to $p$ in~$G.$ We present bounds on $\varphi$ for three classes of graphs: spanning paths, (general and bounded degree) spanning trees, and triangulations. Most of our bounds are tight. To argue this, we describe families of point sets for which no graph from the respective class can achieve a greater incident angle at each vertex.

\paragraph{\bfseries Background.} Our motivation for this research stems from the investigation of pseudo-tri\-an\-gu\-la\-tions, a straight-line framework which---apart from deep combinatorial properties---has \mbox{applications} in motion planning, collision detection, ray shooting and visibility; see~\cite{AABK,HORSSSSSW,KSS,RSS,S} and references therein. Pseudo-triangulations with a minimum number of pseudo-triangles (among all pseudo-triangulations for a given point set) are called \textit{minimum} (or \textit{pointed}) pseudo-triangulations. They can be characterized as plane straight-line graphs where (1) each vertex has an incident angle greater than $\pi$, and (2) the number of edges is maximal, in the sense that the addition of any edge produces an edge-crossing or negates the angle condition.

In this paper, we introduce ``quantified pointedness'' and aim to maximize this parameter: we consider plane straight-line graphs where each vertex has an incident angle of at least $\varphi$---to be maximized. We show that any planar point set admits a triangulation in which each vertex has an incident angle of at least $\frac{2\pi}{3}$. We further consider connected plane straight-line graphs where the number of edges is minimal (spanning trees), and the vertex degree is bounded (spanning trees of bounded degree and spanning paths). Table~\ref{tab:results} lists the obtainable angles of these classes of graphs. Observe that in this context perfect matchings can be described as plane straight-line graphs where each vertex has an incident angle of $2\pi$ and the number of edges is maximal.

\paragraph{\bfseries Related Work.} There is a vast literature on triangulations that are optimal according to certain criteria, see~\cite{AX}. Similar to Delaunay triangulations which maximize the smallest angle over all triangulations for a point set, farthest point Delaunay triangulations minimize the smallest angle over all triangulations for a convex polygon~\cite{E}.  Edelsbrunner~et~al.~\cite{ETW} showed how to construct a triangulation that minimizes the maximum angle among all triangulations for a set of $n$ points in $O(n^2\log n)$ time. If all angles in a triangulation are at least $\frac{\pi}{6}$ then the triangulation contains the relative neighborhood graph as a subgraph~\cite{KW}. The relative neighborhood graph for a point set connects any pair of points which are mutually closest to each other (among all points from the set). 

In applications where small angles have to be avoided by all means, a Delaunay triangulation may not be sufficient in spite of its optimality because even there arbitrarily small angles can occur. By adding so-called Steiner points one can construct a triangulation on a superset of the original points in which there is some absolute lower bound on the size of the smallest angle~\cite{BEG}. Dai~et~al.~\cite{DKC} describe several heuristics to construct minimum weight triangulations (triangulations which minimize the total sum of edge lengths) subject to absolute lower or upper bounds on the occurring angles.

Spanning cycles with angle constraints can be regarded as a variation of the traveling salesman problem. Fekete and Woeginger~\cite{FW} showed that if the cycle may cross itself then any set of at least five points admits a locally convex tour, that is, a tour in which all turns are to the left (or all turns are to the right, respectively). Arkin~et~al.~\cite{AFHMNSS} consider as a measure for (non-) convexity of a point set $S$ the minimum number of (interior) reflex angles (angles $>\pi$) among all plane spanning cycles for $S$, see~\cite{AAK} for recent results. Aggarwal~et~al.~\cite{ACKMS} prove that finding a spanning cycle for a point set which has minimal total angle cost is NP-hard, where the angle cost is defined as the sum of direction changes at the points. Regarding spanning paths, it has been conjectured that each planar point set admits a spanning path with minimum angle at least $\frac{\pi}{6}$~\cite{FW}; recently, a lower bound of $\frac{\pi}{9}$  has been presented~\cite{BPV}.

\begin{wrapfigure}[7]{r}{.31\textwidth}
  \centering
  \vspace{-1.0\baselineskip}
  \includegraphics{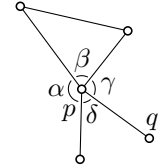}
  \caption{The incident angles of $p$.}
  \label{fig:def}
\end{wrapfigure}
\paragraph{\bfseries Definitions and Notation.} Let $S\subset\R^2$ be a finite set of points in general position, that is, no three points of $S$ are collinear. In this paper we consider plane straight-line graphs $G=(S, E)$ on $S$.  The vertices of $G$ are the points in $S$, the edges of $G$ are straight-line segments that connect two points in $S$, and two edges of $G$ do not intersect except possibly at their endpoints. The \emph{incident angles} of a point $p\in S$ in~$G$ are the angles between any two edges of~$G$ that appear consecutively in the circular order of the edges incident to $p$. We denote the \emph{maximum incident angle} of $p$ in $G$ with $\ma_G(p)$. For a point $p\in S$ of degree at most one we set $\ma_G(p)=2\pi$. We also refer to $\ma_G(p)$ as the \textit{openness} of $p$ in $G$ and call $p\in S$ \textit{$\varphi$-open} in $G$ for some angle $\varphi$ if $\ma_G(p)\ge\varphi$. Consider for example the graph depicted in Fig.~\ref{fig:def}. The point $p$ has four incident edges of $G$ and, therefore, four incident angles. Its openness is $\ma_G(p) = \alpha$. The point $q$ has only one incident angle and correspondingly $\ma_G(q) = 2\pi$.

Similarly we define the \textit{openness} of a plane straight-line graph $G=(S,E)$ as $\ma(G)=\min_{p \in S} \ma_G(p)$ and call $G$ \textit{$\varphi$-open} for some angle $\varphi$ if $\ma(G)\ge\varphi$. In other words, a graph is $\varphi$-open if and only if every vertex has an incident angle of size at least $\varphi$. The \textit{openness} of a class $\mathcal{G}$ of graphs is the supremum over all angles $\varphi$ such that for every finite point set $S\subset\R^2$ in general position there exists a $\varphi$-open connected plane straight-line graph $G$ on $S$ and $G$ is an embedding of some graph from $\mathcal{G}$. For example, the openness of minimum pseudo-triangulations is $\pi$. Without the general position assumption many of these questions become trivial because for a set of collinear points the non-crossing spanning tree is unique---the path that connects them along the line---and its interior points have no incident angle greater than~$\pi$.

The convex hull of a point set $S$ is denoted with $CH(S).$
Points of $S$ on $CH(S)$ are called vertices of $CH(S).$
Let $a$, $b$, and $c$ be three points in the plane that are not
collinear. With $\angle abc$ we denote the counterclockwise angle
between the segment $(b, a)$ and the segment $(b, c)$ at $b$.

\begin{table}[b]
  \centering
  \begin{tabular}{|c|c|c|c|c|}\hline
    Triangulations & Trees & Trees with maxdeg. 3 & Paths (convex sets) & Paths (general) \\\hline
    \rule[-7pt]{0pt}{18pt}$\frac{2\pi}{3}$ & $\frac{5\pi}{3}$ & $\frac{3\pi}{2}$ & $\frac{3\pi}{2}$ & $\frac{5\pi}{4}$\\\hline
  \end{tabular}
  \medskip
  \caption{Openness of several classes of plane straight-line graphs. All given values---except for spanning paths on point sets in general position---are tight.}
  \label{tab:results}
\end{table}
\paragraph{\bfseries Results.}
We study the openness of several classes
of plane straight-line graphs. In particular, in Section~\ref{sec:triang} we give a tight bound of $\frac{2\pi}{3}$ on the openness of triangulations. In Section~\ref{sec:trees} we consider spanning trees, with or without a bound on the maximum vertex degree. For general spanning trees we prove a tight bound of $\frac{5\pi}{3}$; for trees with vertex degree at most three we can still prove a bound of $\frac{3\pi}{2}$, also this bound is tight. Finally, in Section~\ref{sec:paths} we study spanning paths of sets of points in convex or general position. For point sets in convex position we can again show a tight bound of $\frac{3\pi}{2}$; for point sets in general position we prove a non-trivial upper bound of $\frac{5\pi}{4}$. This last bound is not tight, in fact we conjecture that also for point sets in general position the openness of spanning paths is at most $\frac{3\pi}{2}$. Our results are summarized in Table~\ref{tab:results}.

\section{Triangulations}\label{sec:triang}

It is easy to find point sets of any cardinality such that the smallest  angle in any triangulation has to be arbitrary small. In contrast we show that for any point set we can construct a triangulation with a surprisingly large openness.

\begin{figure}[htb]
  \vspace{-1.25\baselineskip}
  \centering
  \includegraphics{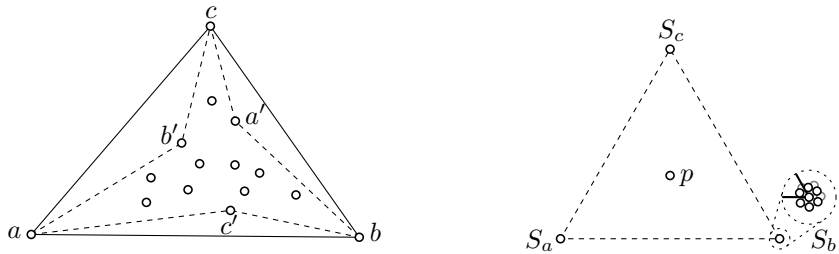}
  \caption{Constructing a $\frac{2\pi}{3}$-open triangulation (left), the openness of triangulations of this point set approaches $\frac{2\pi}{3}$ from above (right).}
  \label{fig:triangle}
  \vspace{-1\baselineskip}
\end{figure}

\begin{theorem}\label{thm:tri}
Triangulations are $\frac{2\pi}{3}$-open and this bound is the best possible.
\end{theorem}
\begin{proof}
  Consider a point set $S\subset{\mathds R}^2$ in general position. Clearly,
  $\mbox{op}_G(p) > \pi$ for every point $p \in \mbox{CH}(S)$ and every plane
  straight-line graph $G$ on $S$. We recursively construct a
  $\frac{2\pi}{3}$-open triangulation $T$ of $S$ by first
  triangulating $\mbox{CH}(S)$; every recursive subproblem consists of a

  Let $S$ be a point set with a triangular convex hull and denote the
  three points of $\mbox{CH}(S)$ with $a$, $b$, and $c$. If $S$ has no
  interior points, then we are done.  Otherwise, let $a'$, $b'$ and
  $c'$ be (not necessarily distinct) interior points of $S$ such that
  the triangles $\Delta a'bc$, $\Delta ab'c$ and $\Delta abc'$ are
  empty (see \figurename~\ref{fig:triangle} (left)). Since the sum of the six
  exterior angles of the hexagon $ba'cb'ac'$ equals $8\pi$, the sum of
  the three angels $\angle ac'b$, $\angle ba'c$, and $\angle cb'a$ is
  at least $2\pi$. In particular, one of them, say $\angle cb'a$, is
  at least $2\pi/3$. We then recurse on the two subsets of $S$ that
  have $\Delta b'bc$ and $\Delta b'ab$ as their respective convex
  hulls.

    The upper bound is attained by a set $S$ of $n$ points as depicted in \figurename~\ref{fig:triangle} (right). $S$ consists of a point~$p$ and of three sets $S_a$, $S_b$, and $S_c$ that each contain $\frac{n-1}{3}$ points. $S_a$, $S_b$, and $S_c$ are placed at the vertices of an equilateral triangle $\Delta$ and $p$ is placed at the barycenter of $\Delta$. Any triangulation $T$ of $S$ must connect~$p$ with at least one point of each of $S_a$, $S_b$, and $S_c$ and hence $\mbox{op}_T(p)$ approaches $\frac{2\pi}{3}$ arbitrarily close from above. 
\end{proof}

\section{Spanning Trees}\label{sec:trees}

In this section we give tight bounds on the $\varphi$-openness of two basic types of spanning trees, namely general spanning trees (Section~\ref{sec:generaltree}) and spanning trees with bounded vertex degree (Section~\ref{sec:boundedtree}). But first we state two technical observations, which will prove useful later.

Consider a point set $S\subset\R^2$ in general position and let $p$ and $q$ be two arbitrary points of $S$. Assume w.l.o.g. that $p$ has smaller $x$-coordinate than $q$. Let $l_p$ and $l_q$ denote the lines through $p$ and $q$ that are perpendicular to the edge $(p, q)$. We define the {\em orthogonal slab} of $(p, q)$ to be the open region bounded by $l_p$ and $l_q.$

\begin{obs}\label{obs:slab}
  Assume that $r \in S \setminus \{p,q\}$ lies in the orthogonal slab
  of $(p, q)$ and above $(p, q)$. Then $\angle qpr \leq \frac{\pi}{2}$
  and $\angle rqp \leq \frac{\pi}{2}$. A symmetric observation holds
  if $r$ lies below $(p, q)$.
\end{obs}

\noindent Recall that the diameter of a point set is the distance between a pair of points that are furthest away from each other.
Let $a$ and $b$ define the diameter of
$S$ and assume w.l.o.g. that $a$ has a smaller $x$-coordinate than
$b$.  Clearly, all points in $S \setminus \{a,b\}$ lie in the
orthogonal slab of $(a, b)$.

\begin{obs}\label{obs:diamter}
  Assume that $r \in S \setminus \{a,b\}$ lies above a diametrical
  segment $(a, b)$ for $S$. Then $\angle arb \geq \frac{\pi}{3}$ and
  hence at least one of the angles $\angle bar$ and $\angle rba$ is at
  most $\frac{\pi}{3}$.  A symmetric observation holds if $r$ lies
  below $(a, b)$.
\end{obs}

\subsection{General Spanning Trees}\label{sec:generaltree}

In this section we consider general spanning trees, that is, spanning trees without any restriction on the degree of their vertices. Throughout this section we use the following notation: we say that an angle $\varphi$ is {\em large} if $\varphi > \frac{\pi}{3}$. Correspondingly, if $\varphi \leq \frac{\pi}{3}$ then we say that $\varphi$ is {\em small}.

\begin{theorem}\label{thm:tree}
Spanning trees are $\frac{5\pi}{3}$-open and this bound is the best possible.
\end{theorem}

\noindent{\it Proof.}
Consider a point set $S\subset\R^2$
in general position and let $a$ and $b$ define the diameter of~$S$.
W.l.o.g. $a$ has a smaller $x$-coordinate than $b$. Let $c \in S
\setminus \{a,b\}$ be the point above $(a, b)$ that is furthest away
from $(a, b)$ and let $d \in S \setminus \{a,b\}$ be the point below
$(a, b)$ that is furthest away from $(a, b)$. (The special case that
$(a, b)$ is an edge of the convex hull of $S$ and hence either~$c$ or~$d$ 
does not exist is handled at the end of the proof.) All points of
$S$ lie within the bounding box defined by the orthogonal slab of $(a,
b)$ and two lines through $c$ and $d$ parallel to $(a, b)$.

\begin{figure}[b]
  \centering
  \includegraphics{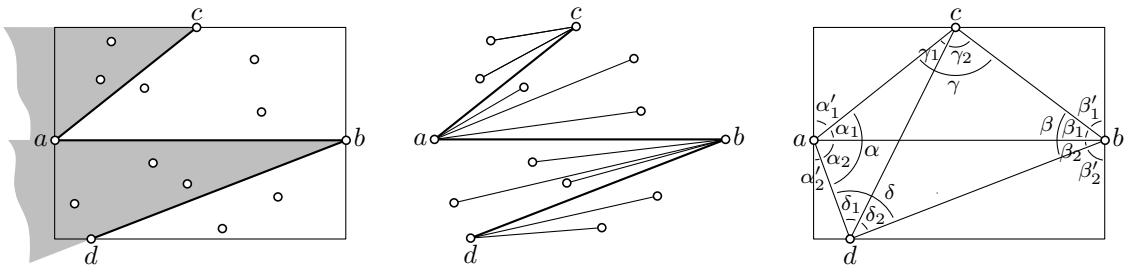}
  \caption{The path $P$ (thick edges) and the cones of $c$ and $b$ (left), the spanning tree constructed from $P$ (middle), the bounding box of $S$ with all relevant angles
  labeled (right).}
  \label{fig:pathdefs}
\end{figure}

To construct a $\frac{5\pi}{3}$-open spanning tree, we first construct
a special $\frac{5\pi}{3}$-open path $P$ whose endpoints are either
$a$ and $b$ or $c$ and $d$. $P$ has the additional property that the
smaller angle at its endpoints between the path and the bounding box
is also small. We extend $P$ to a spanning tree in the following
manner. Every point $p_i$ of $P$ has a small incident angle. Consider
the cone~$C_i$ with apex $p_i$ defined by the edges of $P$ (and the
bounding box if $p_i$ is an endpoint) enclosing the small angle at
$p_i$. When constructing $P$ we ensure that every point $p$ of $S
\setminus P$ is contained in exactly one cone $C_i$. We assemble the
spanning tree by connecting each point in $S \setminus P$ to the apex
of its containing wedge (see Fig.~\ref{fig:pathdefs} (left) and (middle)).

It remains to show that we can always find a path $P$ with the properties described above. We prove this through a case distinction on the size of the angles that are depicted in Fig.~\ref{fig:pathdefs} (right). Since $(a, b)$ is diametrical for $S$,
Observation~\ref{obs:diamter} implies that $\gamma \geq \frac{\pi}{3}$
and $\delta \geq \frac{\pi}{3}$. Furthermore, at least one of
$\alpha_1$ and $\beta_1$ and one of $\alpha_2$ and $\beta_2$ is small.

\begin{description}
   
\item[{\normalfont {\bf Case 1\ }
    Neither at $a$ nor at $b$ both angles ($\alpha_1$ and $\alpha_2$ or $\beta_1$ and $\beta_2$, respectively) are large.}]\hspace{.2\textwidth} 

This means that 
$\alpha_1$ and $\beta_2$ or $\alpha_2$ and $\beta_1$ are small.
If $\alpha_1$ and $\beta_2$ are small, then we choose $P = \langle c,a, b, d \rangle$. $P$ is $\frac{5\pi}{3}$-open and the smaller angles at $c$ and $d$ between~$P$ and the bounding box are at most $\frac{\pi}{3}$. Furthermore, $P$ partitions $S \setminus \{a,b,c,d\}$ into four subsets and each subset is contained in exactly one of the four cones with apex $a$, $b$, $c$, and $d$. Symmetrically, if $\alpha_2$ and $\beta_1$ are small, then $P = \langle c, b, a, d \rangle$.

\item[{\normalfont {\bf Case 2\ }
    Either at $a$ or at $b$ both angles are large.}]\hspace{.2\textwidth}

  W.l.o.g. assume that both $\alpha_1$ and $\alpha_2$ are large and hence
  $\beta_1$ and $\beta_2$ are both small. 
  Futhermore, also all of the angles $\gamma_1$, $\delta_1$, 
  $\alpha_1' = \frac{\pi}{2}-\alpha_1$,
  and $\alpha_2' = \frac{\pi}{2} - \alpha_2$  are small.

\begin{figure}
  \centering
  \includegraphics{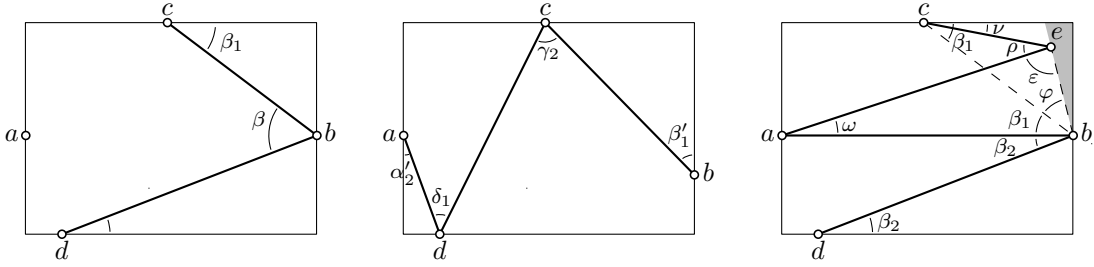}
  \caption{Case 2.1 (left), Case 2.2.1 for $\gamma_2$ small and $\beta_1'$ large (middle), Case 2.2.2.2  for $\gamma_2$ small and $\beta_1'$ large (right).}
  \label{fig:case2}
  \vspace{-.5\baselineskip}
\end{figure}

\item[{\normalfont {\bf Case 2.1\ }
    $\beta = \beta_1 + \beta_2$ is small.}]\hspace{.2\textwidth}

We choose $P = \langle c, b, d \rangle$ (see Fig.~\ref{fig:case2} (left)). $P$ is $\frac{5\pi}{3}$-open and the smaller angles at $c$ and $d$ between $P$ and the bounding box are at most $\frac{\pi}{3}$. $P$ partitions $S \setminus \{b,c,d\}$ into three subsets and each subset is contained in exactly one of the three cones with apex $b$, $c$, and $d$.

\item[{\normalfont {\bf Case 2.2\ }
    $\beta = \beta_1 + \beta_2$ is large.}]\hspace{.2\textwidth}
 
Since $\beta = \beta_1 + \beta_2$ is large it follows that at least one of $\gamma_2$ and $\delta_2$ 
and at least one of $\beta_1' = \frac{\pi}{2}-\beta_1$ and $\beta_2' = \frac{\pi}{2}-\beta_2$ is small.

\item[{\normalfont {\bf Case 2.2.1\ } Both $\beta_1'$ and
    $\gamma_2$ are small or both $\beta_2'$ and $\delta_2$ are
    small.}]\hspace{.2\textwidth}

  If both $\beta_1'$ and $\gamma_2$ are small then we choose $P =
  \langle a, d, c, b \rangle$ (see Fig.~\ref{fig:case2} (middle)). $P$ is
  $\frac{5\pi}{3}$-open and partitions $S \setminus \{a,b,c,d\}$ into
  four subsets which each are contained in exactly one of the four
  cones with apex $a$, $b$, $c$, and $d$. Symmetrically, if both
  $\beta_2'$ and $\delta_2$ are small, then we can use the path 
  $P = \langle a, c, d, b \rangle$.

\item[{\normalfont {\bf Case 2.2.2\ }
    Neither both $\beta_1'$ and $\gamma_2$ are small nor both $\beta_2'$ and $\delta_2$ are small.}]\hspace{.2\textwidth}

    Consider the subset $S_c$ of $S$ that consists of the points above $(c, b)$, and 
    the subset $S_d$ of $S$ that consists of the points below $(d, b)$.

\item[{\normalfont {\bf Case 2.2.2.1\ }
    $\beta_1'$ and $\delta_2$ are large. Thus $\gamma_2$ is small.}]\hspace{.2\textwidth}

\item[{\normalfont {\bf Case 2.2.2.1.1\ }
    $\angle pbc$ is small for all points $p \in S_c$.}]\hspace{.2\textwidth}

    All edges from $b$ to the points in $S_c \cup \{c\}$ 
    lie in a wedge with angle $\tilde{\beta}_1'$ smaller than $\frac{\pi}{3}$.
    As $\gamma_2$ is small and $\tilde{\beta}_1'$ replaces $\beta_1'$ we choose 
    $P = \langle a, d, c, b \rangle$ as in Case 2.2.1 (see Fig.~\ref{fig:case2} (middle)).
    
\item[{\normalfont {\bf Case 2.2.2.1.2\ }
    $\angle pbc$ is large for at least one point $p \in S_c$.}]\hspace{.2\textwidth}

    Let $e \in S_c$ be the point such that $\varphi = \angle
    ebc$ is largest among the points in $S_c$. We choose $P = \langle c,
    e, a, b, d \rangle$ (see Fig.~\ref{fig:case2} (right)). The angle
    $\nu$ is small since it is smaller than $\beta_1$, and $\beta_1$ is
    small. Furthermore, $\varphi$ is large by definition of $e$ and
    Observation~\ref{obs:diamter} implies that $\angle aeb =
    \varepsilon$ is at least $\frac{\pi}{3}$. Summing the angles within
    $\triangle cbe$ yields $\varrho+\beta_1-\nu +\varphi+\varepsilon=\pi$. 
    Therefore $\varrho+\beta_1-\nu$ is small, and as $\beta_1-\nu \geq 0$, 
    also $\varrho$ is small. Similarly, the angle sum
    within $\triangle abe$ is $\omega+\beta_1+\varphi+\varepsilon=\pi$
    and therefore $\omega$ is small. In summary, all of
    $\beta_2$, $\omega$, $\varrho$, and $\nu$ are small and hence $P$ is
    $\frac{5\pi}{3}$-open.
    $P$ partitions $S \setminus \{a,b,c,d,e\}$ into five subsets, and 
    since the gray-shaded region in Fig.~\ref{fig:case2} (right) does not contain any points of $S$ by choice of $e$, each subset is contained in exactly one of the five
    cones with apex $a$, $b$, $c$, $d$, and $e$.

\item[{\normalfont {\bf Case 2.2.2.2\ }
    $\beta_2'$ and $\gamma_2$ are large. Thus $\delta_2$ is small.}]\hspace{.2\textwidth}

    If $\angle dbq$ is small for all points $q \in S_d$ then, 
    using symmetric arguments as in Case~2.2.2.1.1, 
    we choose $P = \langle a, c, d, b \rangle$ like in Case~2.2.1.
    
    If $\angle dbq$ is large for at least one point $q \in S_d$, 
    then let $f \in S_d$ be the point maximizing this angle. 
    Then, using symmetric arguments as in Case~2.2.2.1.2, 
    we choose $P = \langle c, b, a, f, d \rangle$.

\end{description}

\noindent
Finally, if $(a, b)$ is an edge of the convex hull then either $c$ or
$d$ does not exist. If $c$ does not exist, then we can choose either
$P = \langle a, b, d \rangle$ or $P = \langle b, a, d \rangle$. A
symmetric argument holds if $d$ does not exist.
 
\begin{wrapfigure}[10]{r}{.5\textwidth}
  \centering
  \includegraphics{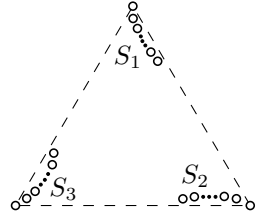}
  \caption{Every spanning tree of $S$ is at most $\frac{5\pi}{3}$-open.}
  \label{fig:uppertree}
\end{wrapfigure}
The upper bound is attained by the point set depicted in
Fig.~\ref{fig:uppertree}.  Each of the sets $S_i, i \in
{1,2,3}$ consists of $\frac{n}{3}$ points.  If a point $p \in S_1$ is
connected to any other point from $S_1 \cup S_2$, then it can only be
connected to a point of $S_3$ forming an angle of at least
$\frac{\pi}{3} - \varepsilon$. As the same argument holds for $S_2$
and $S_3$, respectively, any connected graph, and thus any spanning
tree on~$S$ is at most $\frac{5\pi}{3}$-open.%
\hfill\qed

\subsection{Spanning~Trees~of~Bounded~Vertex~Degree}\label{sec:boundedtree}

By construction, the spanning trees obtained in the previous section might have arbitrarily large vertex degree which can be undesirable. Hence in the following we consider spanning trees with bounded maximum vertex degree and derive tight bounds on their openness.

\begin{theorem}\label{thm:tree_delta}
  Let $S\subset\R^2$ be a set of $n$ points in general position. There
  exists a $\frac{3\pi}{2}$-open spanning tree $T$ of $S$ such that
  every point from $S$ has vertex degree at most three in $T$. The
  angle bound is best possible, even for the much broader class of
  spanning trees of vertex degree at most $n-2$.
\end{theorem}

\noindent{\it Proof.}
  We show that $S$ has a $\frac{3\pi}{2}$-open spanning tree
  with maximum vertex degree three. To do so, we first describe a
  recursive construction that results in a $\frac{3\pi}{2}$-open
  spanning tree with maximum vertex degree four. We then refine our
  construction to yield a spanning tree of maximum vertex degree
  three.

\begin{figure}[b]
  \centering
  \includegraphics{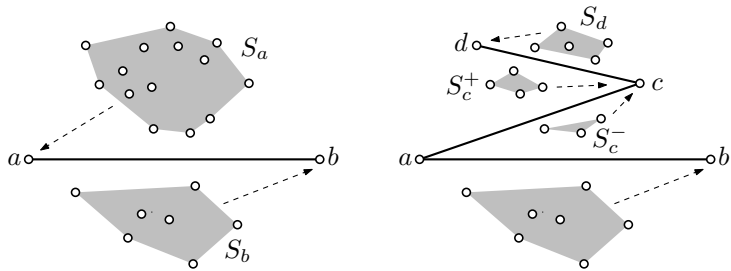}
  \caption{Constructing a $\frac{3\pi}{2}$-open spanning tree with maximum vertex degree four.}
\label{fig:treedeg4}
\end{figure}
Let $a$ and $b$ define the diameter of $S$. W.l.o.g. $a$ has a   smaller $x$-coordinate than $b$. The edge $(a, b)$ partitions   $S\setminus\{a,b\}$ into two (possibly empty) subsets: the set $S_a$   of the points above $(a, b)$ and the set $S_b$ of the points below   $(a, b)$. We assign $S_a$ to $a$ and $S_b$ to $b$ (see   Fig.~\ref{fig:treedeg4}). Since all points of $S \setminus \{a,b\}$ lie in the orthogonal slab of $(a, b)$ we can connect any point $p   \in S_a$ to $a$ and any point $ q \in S_b$ to $b$ and by this   obtain a $\frac{3\pi}{2}$-open path $P = \langle p, a, b, q   \rangle$. Based on this observation we recursively construct a spanning tree of vertex degree at most four.

  If $S_a$ is empty, then we proceed with $S_b$. If $S_a$ contains
  only one point~$p$ then we connect $p$ to~$a$. Otherwise consider a
  diametrical segment $(c, d)$ for $S_a$. W.l.o.g. $d$ has a smaller
  \mbox{$x$-coordinate} than $c$ and~$d$ lies above $(a, c)$. Either $\angle
  adc$ or $\angle dca$ must be less than $\frac{\pi}{2}$. W.l.o.g.
  assume that $\angle dca < \frac{\pi}{2}$. Hence we can connect $d$
  via $c$ to $a$ and obtain a $\frac{3\pi}{2}$-open path \mbox{$P = \langle d, c, a, b \rangle$}. 
  The edge $(d, c)$ partitions $S_a$ into two
  (possibly empty) subsets: the set $S_d$ of the points above $(d, c)$
  and the set $S_c$ of the points below $(d, c)$. The set $S_c$ is
  again partitioned by the edge $(a, c)$ into a set $S^+_c$ of points
  that lie above $(a, c)$ and a set $S^-_c$ of points that lie below
  $(a, c)$. We assign $S_d$ to $d$ and both $S^+_c$ and $S^-_c$ to $c$
  and proceed recursively. 
  (Note that by Observation~\ref{obs:slab} $\angle dce \le \frac{\pi}{2} \ \forall e \in S_c$, 
  and $\angle cde \le \frac{\pi}{2}  \ \forall e \in S_d$.)

  The algorithm maintains the following two invariants: $(i)$ at most
  two sets are assigned to any point of $S$, and $(ii)$ if a set $S_p$
  is assigned to a point $p$ then $p$ can be connected to any point of
  $S_p$ and $\ma_T(p) \geq \frac{3\pi}{2}$ for any resulting tree $T$.

\begin{figure}[t]
    \centering
    \includegraphics{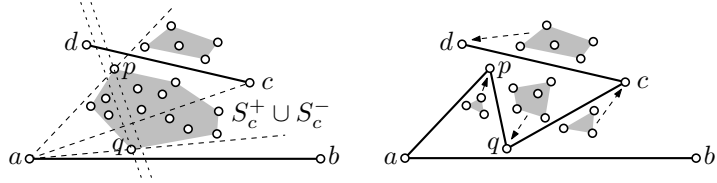}
    \caption{Constructing a $\frac{3\pi}{2}$-open spanning tree with maximum vertex degree three.}
    \label{fig:treedeg3}
\end{figure}

We now refine our construction to obtain a $\frac{3\pi}{2}$-open
spanning tree of maxi\-mum vertex degree three. If $S^+_c$ is empty then we assign $S^-_c$ to $c$, and vice versa. Otherwise, consider the tangents
from $a$ to $S_c$ and denote the points of tangency with $p$ and $q$
(see Fig.~\ref{fig:treedeg3}). Let $l_p$ and $l_q$ denote the
lines through $p$ and $q$ that are perpendicular to $(a, c)$. W.l.o.g.
$l_q$ is closer to $a$ than $l_p$. We replace the edge $(a, c)$ by the
three edges $(a, p)$, $(p, q)$, and $(q, c)$. The resulting path is
\mbox{$\frac{3\pi}{2}$-open} and partitions $S_c$ into three sets which can
be assigned to $p$, $q$, and $c$ while maintaining invariant~$(ii)$.
The refined recursive construction assigns at most one set to every
point of $S$ and hence constructs a $\frac{3\pi}{2}$-open spanning
tree with maximum vertex degree three.

\begin{wrapfigure}[8]{l}{.4\textwidth}
    \centering
    \vspace{-.5\baselineskip}
    \includegraphics{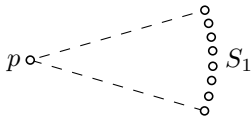}
    \caption{Every spanning tree of this point set with vertex degree at most $n-2$ is at most $\frac{3\pi}{2}$-open.}
    \label{fig:treedeg3tight}
\end{wrapfigure}
The upper bound is attained by a set $S$ of $n$ points as depicted in Fig.~\ref{fig:treedeg3tight}. $S$~consists of a subset~$S_1$ of~$n-1$ near-collinear points close together and one point $p$ far away. In order to construct any connected graph with maximum degree at most $n-2$, one point of $S_1$ has to be connected to another point of $S_1$ and to $p$. Thus any spanning tree on $S$ with maximum degree at most $n-2$ is at most $\frac{3\pi}{2}$-open. 
\hfill\qed

\section{Spanning Paths}\label{sec:paths}

Spanning paths can be regarded as spanning trees with maximum vertex
degree two. Therefore, the upper bound construction in 
Fig.~\ref{fig:treedeg3tight} applies to spanning paths as well. We 
show in Section~\ref{sec:convex} below that the resulting bound of $\frac{3\pi}{2}$ is tight for points in convex position, even in a very strong sense: there exists a
$\frac{3\pi}{2}$-open spanning path starting from any predefined point. For points in general position we prove a non-trivial upper bound of $\frac{5\pi}{4}$ in Section~\ref{sec:path_general}.

\subsection{Point Sets in Convex Position}\label{sec:convex}

\begin{wrapfigure}[8]{r}{.21\textwidth}
  \centering
  \vspace{-.5\baselineskip}
  \includegraphics{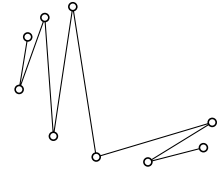}
  \caption{A zigzag path.}
  \label{fig:path_convex_zigzag}
\end{wrapfigure}

Consider a set $S\subset\R^2$ of $n$ points in convex position. We can
construct a spanning path for~$S$ by starting at an arbitrary point
$p\in S$ and recursively taking one of the tangents from $p$ to~$\CH(S\setminus\{p\})$. 
As long as $|S|>2$, there are two tangents
from $p$ to $\CH(S\setminus\{p\})$: the left tangent is the oriented
line $t_\ell$ through $p$ and a point  $p_{\ell}\in
S\setminus\{p\}$ (oriented in direction from $p$ to $p_{\ell}$) such
that no point from $S$ is to the left of $t_\ell$. Similarly, the
right tangent is the oriented line $t_r$ through $p$ and a point
$p_r\in S\setminus\{p\}$ (oriented in direction from $p$ to $p_{r}$)
such that no point from $S$ is to the right of $t_r$. If we take the
left and the right tangent alternatingly, see
Fig.~\ref{fig:path_convex_zigzag}, we
call the resulting path a\textit{zigzag} path for~$S$.%

\begin{theorem}\label{thm:path}
  Every finite point set in convex position in the plane admits a
  spanning path that is $\frac{3\pi}{2}$-open and this bound is the best
  possible.
\end{theorem}

We present two different proofs for this theorem. First an
existential proof using counting arguments and then a constructive
proof that, in addition, provides a stronger claim. To see that the bound of $\frac{3\pi}{2}$ is tight, consider again the point set in Fig.~\ref{fig:treedeg3tight}. 
\begin{proof}[Theorem~\ref{thm:path}, existential]
  As a zigzag path is completely determined by one of its endpoints
  and the direction of the incident edge, there are exactly $n$ zigzag
  paths for $S$. (Count directed zigzag paths: there are $n$ choices
  for the starting point and two possible directions to continue, that is, $2n$ directed zigzag paths and, therefore, $n$ (undirected) zigzag paths.)

  Now consider a point $p\in S$ and sort all other points of~$S$
  radially around~$p$, starting with one of the neighbors of~$p$ along
  $\CH(S)$. Any angle that occurs at~$p$ in some zigzag path for~$S$
  is spanned by two points that are consecutive in this radial order.
  Moreover, any such angle occurs in exactly one zigzag path because
  it determines the zigzag path completely. Since the sum of all these
  angles at~$p$ is less than~$\pi$, for each point~$p$ at most one
  angle can be $\ge\frac{\pi}{2}$. Furthermore, if~$p$ is an endpoint
  of a diametrical segment for $S$ then all angles at~$p$ are
  $<\frac{\pi}{2}$. Since there is at least one diametrical segment
  for $S$, there are at most $n-2$ angles $>\frac{\pi}{2}$ in all
  zigzag paths together. Thus, there exist at least two spanning
  zigzag paths that have no angle $>\frac{\pi}{2}$, that is, they are
  $\frac{3\pi}{2}$-open.    
\end{proof}

\begin{wrapfigure}[8]{r}{.28\textwidth}
  \centering
  \vspace{-0.75\baselineskip}
  \includegraphics{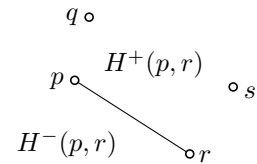}
  \caption{$(p,r)$ is expanding in direction ``+''.}
  \label{fig:expand}
\end{wrapfigure}
Before we present the constructive proof, we give some technical definitions and observations. For two distinct points $p,r\in\R^2$ denote by $H^-(p,r)$ the set of points on or to the right of the ray $\overrightharpoonup{pr}$, that is, those $t\in\R^2$ for which $\angle prt\le\pi$. Correspondingly, denote by $H^+(p,r)$ the set of points on or to the left of the ray $\overrightharpoonup{pr}$, that is, those $t\in\R^2$ for which $\angle prt\ge\pi$, see Fig.~\ref{fig:expand}. Let $S^+(p,r):=S\cap H^+(p,r)$ and $S^-(p,r):=S\cap H^-(p,r)$. Consider a directed segment $(p,r)$, for some $p,r\in S$, and a direction $\tau\in\{+,-\}$. Denote by $q$ and $s$ the neighbors of~$p$ and~$r$, respectively, along $\CH(S)$ that are in $S^{\tau}(p,r)$ (possibly, $q=s$ or even $q=r$ and $s=p$). We call $(p,r)$ \textit{expanding} in direction $\tau$ if the two rays $\overrightharpoonup{qp}$ and~$\overrightharpoonup{sr}$ intersect outside $H^{\tau}(p,r)$; otherwise, $(p,r)$ is called \textit{non-expanding} in direction $\tau$. Observe that if $|S^{\tau}(p,r)|\le 3$ then $(p,r)$ is non-expanding in direction $\tau$.

\begin{proof}[Theorem~\ref{thm:path}, constructive]
  The proof uses the following more general claim.
  \begin{myclaim}\label{claim:rays}
  Consider a directed segment $(p,r)$, for some $p,r\in S$,
  and a direction $\tau\in\{+,-\}$. Denote by $q$ and $s$ the neighbors of
    $p$ and $r$ (resp.) along $\CH(S)$ that are in
    $S^{\tau}(p,r)$ (possibly, $q=s$ or even $q=r$ and $s=p$).
    Suppose that $(p,r)$ is non-expanding in direction $\tau$ and that
    \begin{itemize} 

    \item \quad if \quad $\tau={+}$ \quad then \quad $\angle trp\le\frac{\pi}{2}$ \quad for all \quad 
      $t\in S^+(p,r)\setminus\{p,r\}$;

    \item \quad if \quad $\tau={-}$ \quad then \quad  $\angle prt\le\frac{\pi}{2}$ \quad for all \quad 
      $t\in S^-(p,r)\setminus\{p,r\}$.

    \end{itemize}
    Then there is a $\frac{3\pi}{2}$-open spanning path for
    $S^{\tau}(p,r)$ that starts with $\langle p,r\rangle$.
  \end{myclaim}

  The condition on the angles above states 
  that $\langle p,r\rangle$ can be extended to a $\frac{3\pi}{2}$-open
  path by any single point from $S^{\tau}(p,r)\setminus\{p,r\}$. Specifically, all conditions of the claim are fulfilled by any
  diametrical segment $(p,r)$ of $S$, for both of its two possible
  orientations. Hence, applying the claim to both $(p,r)$ and
  direction ``$+$'' as well as $(r,p)$ and direction ``$+$'' yields
  Theorem~\ref{thm:path}.%
\end{proof}
  
It remains to prove Claim~\ref{claim:rays}.

\begin{proof}[Claim~\ref{claim:rays}] We use induction on $|S^{\tau}(p,r)|$. The statement is trivial if
  $|S^{\tau}(p,r)|\in\{2,3\}$. Therefore let $|S^{\tau}(p,r)|\ge 4$
  and consider the segment $(q,s)$. Observe that by convexity of $S$
  the segment $(q,s)$ is non-expanding in direction $\tau$ and
  $S^{\tau}(q,s)=S^{\tau}(p,r)\setminus\{p,r\}$. From now on, assume
  that $\tau={+}$; the case $\tau={-}$ is symmetric.
  
  \begin{figure}[h]
      \vspace{-.5\baselineskip}
      \centering%
      \subfigure[Case 1: Angles.]{\label{fig:convex:1a}%
       \includegraphics{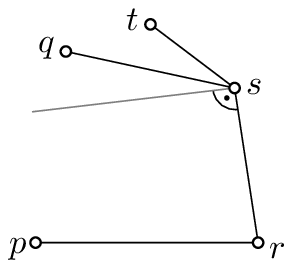}}
      \hfil%
      \subfigure[Case 1: Path.]{\label{fig:convex:1b}%
        \includegraphics{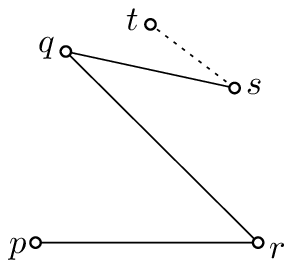}}
      \hfil%
      \subfigure[Case 2: Angles.]{\label{fig:convex:2a}%
        \includegraphics{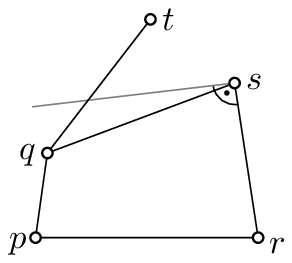}}
      \hfil%
      \subfigure[Case 2: Path.]{\label{fig:convex:2b}%
        \includegraphics{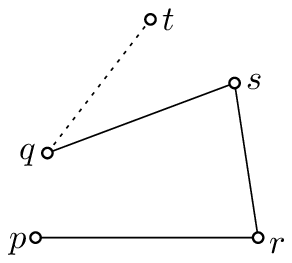}}
      \hfil%
      \caption{\label{fig:convex:1}\label{fig:convex:2} Constructing a
        $\frac{3\pi}{2}$-open spanning path.}
      \vspace{-.5\baselineskip}
\end{figure} 

  \begin{description}
  \item[{\normalfont {\bf Case 1\ } $\angle qsr\ge\frac{\pi}{2}$}.]
    \hspace{.2\textwidth}

    Illustrated in Fig.~\ref{fig:convex:1}(a) and \ref{fig:convex:1}(b)  ---  $(q,s)$
    fulfills the angle condition, since for every \mbox{$t\in S^+(q,s)\setminus\{q,s\}$}
    \[
    \angle tsq=\angle tsr-\angle qsr\le\angle tsr-\frac{\pi}{2},
    \]
    and since $\angle tsr\le\pi$ by convexity of~$S$. Thus, we can extend
    $\langle q,s\rangle$ to a $\frac{3\pi}{2}$-open spanning path \mbox{for $S^+(q,s)$} inductively. That path together with $\langle
    p,r,q\rangle$ forms a $\frac{3\pi}{2}$-open spanning path for~$S$.

  \item[{\normalfont {\bf Case 2\ } $\angle
      qsr<\frac{\pi}{2}.$}]\hspace{.2\textwidth}

    Illustrated in Fig.~\ref{fig:convex:2}(c) and \ref{fig:convex:2}(d)  ---  as $(p,r)$ is
    non-expanding in direction ``$+$'', we have $\angle srp+\angle
    rpq\le\pi$. Summing the angles within the quadrilateral
    $(p,r,s,q)$ yields
    \[
    2\pi=\angle srp+\angle rpq+\angle pqs+\angle
    qsr<\frac{3\pi}{2}+\angle pqs\,,
    \]
    that is, $\angle pqs>\frac{\pi}{2}$. We conclude that for every
    $t\in S^-(s,q)\setminus\{q,s\}$
    \[
    \angle sqt=\angle pqt-\angle pqs<\angle
    pqt-\frac{\pi}{2}\le\frac{\pi}{2}
    \]
    as $\angle pqt\le\pi$ by convexity of $S$. Thus, we can extend
    $\langle s,q \rangle$ to a $\frac{3\pi}{2}$-open spanning path for
    $S^-(s,q)$ inductively. That path together with $\langle
    p,r,s\rangle$ forms a $\frac{3\pi}{2}$-open spanning path for $S$. 
  \end{description}%
\end{proof}

In the remainder of this section we prove a statement that is even stronger than Theorem~\ref{thm:path}: for points in convex position there exists a
$\frac{3\pi}{2}$-open spanning path starting at any arbitrary point.

\begin{corollary}
    For any finite set $S\subset\R^2$ of points in convex position and
  any $p\in S$ there exists a $\frac{3\pi}{2}$-open spanning path
  for $S$ which has $p$ as an endpoint.
\end{corollary}

\begin{wrapfigure}[7]{r}{.4\textwidth}
  \centering
  \includegraphics{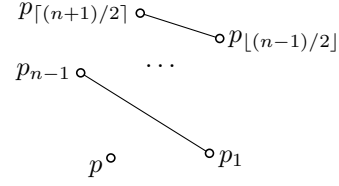}
  \caption{Segments ``parallel to'' $p$.}
  \label{fig:cor5}
\end{wrapfigure}
\noindent\textit{Proof.} For $|S|\le 3$ the statement is trivial. Hence suppose
$|S|\ge 4$. Denote by $(p=p_0,p_1,\ldots,p_{n-1})$ the sequence of
points along $\CH(S)$ in counterclockwise order and consider the
sequence
  \[
  (s_i=(p_i,p_{n-i}))_{i=1\ldots\lfloor(n-1)/2\rfloor}
  \]
  of segments ``parallel to $p$'', as depicted in
  Fig.~\ref{fig:cor5}. \mbox{Observe} that $s_{\lfloor(n-1)/2\rfloor}$
  is non-expanding in direction~``$-$'' because there are no more than
  three points in
  $S^-(p_{\lfloor(n-1)/2\rfloor},p_{\lceil(n+1)/2\rceil})$.
  Analogously, $s_1$ is non-expanding in direction ``$+$''.
  Therefore, the minimum index $k$, $1\le k\le\lfloor(n-1)/2\rfloor$,
  for which $s_k$ is non-expanding in direction ``$-$'' is well
  defined.

  If $k=1$ then $s_1$ is a segment that is non-expanding for both
  directions. Otherwise, by the minimality of $k$ the segment
  $s_{k-1}$ is expanding for direction ``$-$''. By definition, if $s_i$
  is expanding in direction ``$-$'' then $s_{i+1}$ is non-expanding in
  direction ``$+$'', for \mbox{$1\le i<\lfloor(n-1)/2\rfloor$}. Thus, in any
  case, $s_k$ is a segment that is non-expanding for both directions.

  Suppose there is a point $q\in
  S^-(p_k,p_{n-k})\setminus\{p_k,p_{n-k}\}$ for which $\angle
  p_kp_{n-k}q>\frac{\pi}{2}$. Then the convexity of $S$ implies
  $\angle rp_{n-k}p_k<\frac{\pi}{2}$ for all $r\in
  S^+(p_k,p_{n-k})\setminus\{p_k,p_{n-k}\}$.  Moreover, as $s_k$ is
  non-expanding in direction ``$-$'' we have $\angle
  rp_kp_{n-k}<\frac{\pi}{2}$. Application of Claim~\ref{claim:rays} to
  $(p_k,p_{n-k})$ and $\tau=+$ yields a $\frac{3\pi}{2}$-open spanning
  path for $S^+(p_k,p_{n-k})$ starting with $\langle
  p_k,p_{n-k}\rangle$. Similarly, applying Claim~\ref{claim:rays} to
  $(p_{n-k},p_k)$ and $\tau=+$ we obtain a $\frac{3\pi}{2}$-open
  spanning path for $S^+(p_{n-k},p_k)$ starting with $\langle
  p_{n-k},p_k\rangle$. Combining both paths provides the desired
  $\frac{3\pi}{2}$-open spanning path for~$S$. This path has $p$ as
  one of its endpoints by construction.

  In a symmetric way, we can handle the case that there is a point
  $s\in S^+(p_k,p_{n-k})\setminus\{p_k,p_{n-k}\}$ for which $\angle
  p_{n-k}p_ks>\frac{\pi}{2}$. Finally, if neither of the points $q$
  and $s$ exist, we can apply Claim~\ref{claim:rays} to
  $(p_k,p_{n-k})$ and $\tau=-$ as well as to $(p_{n-k},p_k)$ and
  $\tau=-$ and in this way obtain a $\frac{3\pi}{2}$-open spanning
  path for $S$ which has $p$ as one of its endpoints.
  \hfill\qed

\subsection{General Point Sets}\label{sec:path_general}

We finally consider the openness of spanning paths for general point sets. Unfortunately we cannot give tight bounds in this case, but we do present a non-trivial upper bound on the openness. Let $S\subset{\mathds R}^2$ be a set of $n$ points in general
position.  For a suitable labeling of the points of $S$ we denote a
spanning path for (a subset of $k$ points of) $S$ with
$\langle p_1,\ldots,p_k \rangle$, where we call $p_1$ the starting point of
the path.  Now Lemma~\ref{thm:5pi4thpath_simple} follows directly from Theorem~\ref{thm:5pi4thpath}.

\begin{lemma}\label{thm:5pi4thpath_simple}
Spanning paths are $\frac{5\pi}{4}$-open.
\end{lemma}

\begin{theorem}\label{thm:5pi4thpath}
Let $S$ be a finite point set in general position in the plane. Then
\begin{itemize}
\item[(1)] For every vertex $q$ of the convex hull of $S$, there
exists a \mbox{$\frac{5\pi}{4}$-open} spanning path
$\langle q,p_1,\ldots,p_k \rangle$ on $S$ starting at $q$.

\item[(2)] For every edge $\overline{q_1q_2}$ of the convex hull of
$S$ there exists a \mbox{$\frac{5\pi}{4}$-open} spanning path starting
at either $q_1$ or $q_2$ and using the edge $\overline{q_1q_2}$, that is, a spanning
path $\langle q_1,q_2,p_1,\ldots,p_k\rangle$ or $\langle q_2,q_1,p_1,\ldots,p_k\rangle$.
\end{itemize}
\end{theorem}

\begin{proof}
For each vertex $p$ in a path $G$ the maximum incident angle $\mbox{op}_G(p)$
is the larger of the two incident angles (except for start- and endpoint of the path).
To simplify the discussion we consider the smaller angle at each
point and prove that we can construct a spanning path such that this angle
is at most $\frac{3\pi}{4}$. We denote with $(q,S)$ a spanning path for $S$ starting at $q$, and with $(\overline{q_{1}q_{2}},S)$ a spanning path for $S$ starting with the edge connecting $q_1$ and $q_2.$ The \emph{outer normal cone} of a vertex $y$ of a convex polygon is the region between two half-lines that start at $y$, are respectively perpendicular to the two edges incident at $y$, and are both in the exterior of the polygon. We prove part (1) and (2) of
Theorem~\ref{thm:5pi4thpath} by induction on $|S|$. The base
cases for $|S|=3$ clearly hold.

\medskip

\noindent\textbf{Induction for (1):} Let ${\cal K} = CH(S\setminus\{q\}).$

\begin{description}

\item[{\normalfont {\bf Case 1.1\ }
    \raisebox{-.485\baselineskip}{\parbox{.85\textwidth}{$q$ lies between the outer normal cones of two consecutive vertices $y$ and
$z$ of $\cal K$, where $z$ lies to the right of the ray $\overrightharpoonup{qy}.$}}}]\hspace{.2\textwidth}

Induction on $(\overline{yz},S\setminus\{q\})$ yields a \mbox{$\frac{5\pi}{4}$-open} 
spanning path $\langle y,z,p_1,\ldots,p_k\rangle$ or  
$\langle z,y,p_1,\ldots,p_k\rangle$ of $S\setminus\{q\}$.
Obviously $\angle qyz \leq \frac{\pi}{2} < \frac{3\pi}{4}$
and $\angle yzq \leq \frac{\pi}{2} < \frac{3\pi}{4}$, and thus we get a
\mbox{$\frac{5\pi}{4}$-open} spanning path $\langle q,y,z,p_1,\ldots,p_k\rangle$
or $\langle q,z,y,p_1,\ldots,p_k\rangle$ for $S$ (see Fig.~\ref{fig:pathCases1} (left)).

\item[{\normalfont {\bf Case 1.2\ }
    $q$ lies in the outer normal cone of a vertex of $\cal K$.}]\hspace{.2\textwidth}

Let $p$ be that vertex and
let $y$ and $z$ be the two vertices of $\cal K$ adjacent to $p,$ $z$ being to the right of the ray $\overrightharpoonup{py}.$
The three angles $\angle qpz$, $\angle zpy$ and $\angle ypq$ around $p$ obviously add up to $2\pi$. We consider subcases according to which of the three angles is the smallest, the cases of $\angle qpz$ and $\angle ypq$ being symmetric (see Fig.~\ref{fig:pathCases1} (middle)).

\item[{\normalfont {\bf Case 1.2.1\ }
    $\angle zpy$ is the smallest of the three angles.}]\hspace{.2\textwidth}

Then, in particular, $\angle zpy < \frac{3\pi}{4}$. Assume without loss of generality that $\angle qpz$ is smaller than~$\angle ypq$ and, in particular,
that it is smaller than $\pi$. Since $q$ is in the normal cone of $p,$  $\angle qpz$ is at least $\frac{\pi}{2}$,
hence $\angle pzq$ is at most $\frac{\pi}{2} < \frac{3\pi}{4}$. Let $S'=S\setminus \{q,z\}$ and consider the path
that starts with $q$ and $z$ followed by $(p,S')$, that is $\langle q,z,p,p_1,\ldots,p_k\rangle.$
Note that $\angle zpp_1 \leq \angle zpy.$

\item[{\normalfont {\bf Case 1.2.2\ }
    $\angle ypq$ is the smallest of the three angles.}]\hspace{.2\textwidth}

Then $\angle ypq < \frac{3\pi}{4}.$
Moreover, in this case all three angles $\angle qpz$, $\angle ypq$ and $\angle zpy$ are at least $\frac{\pi}{2}$, the first two because $q$ lies in the normal cone of $p$, the latter because it is not the smallest of the three angles.
We have $\angle qyp < \frac{\pi}{2}$ because this angle lies in the triangle containing $\angle ypq \geq \frac{\pi}{2}$,
and $\angle ypq < \frac{3\pi}{4}$ by assumption.
We iterate on $(\overline{py},S\setminus\{q\})$  and get a
\mbox{$\frac{5\pi}{4}$-open} spanning path on $S\setminus\{q\}$ by
induction, which can be extended to a \mbox{$\frac{5\pi}{4}$-open}
spanning path on $S,$ $\langle q,p,y,p_1,\ldots,p_k\rangle$
or $\langle q,y,p,p_1,\ldots,p_k\rangle$, respectively.
\end{description}

\begin{figure}[b]
  \centering
  \includegraphics{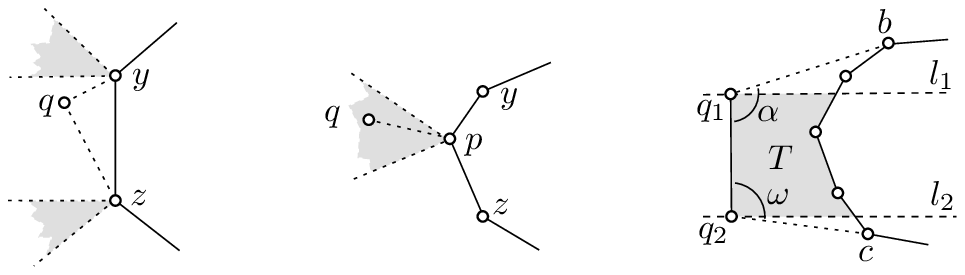}
  \caption{Case 1.1 (left), Case 1.2 (middle), Case 2 (right).}
  \label{fig:pathCases1}
\end{figure}

\begin{figure}[t]
  \centering
  \includegraphics{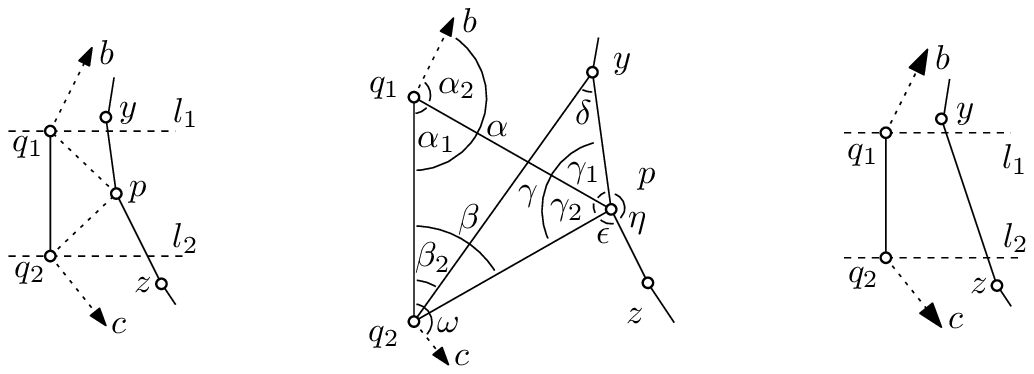}
  \caption{Case 2.2.1 (left), Case 2.2.1.[1$|$2] (middle), Case 2.2.2 (right).}
  \label{fig:pathCases2}
  \vspace{-.2\baselineskip}
\end{figure}

\noindent\textbf{Induction for (2):}
Let $b$ and $c$ be the neighboring vertices of $q_1$ and $q_2$ on
$CH(S)$, such that $CH(S)$ reads $\ldots,b,q_1,q_2,c,\ldots$ in
ccw order (see Fig.~\ref{fig:pathCases1} (right)).

\begin{description}

\item[{\normalfont {\bf Case 2.1\ }
    $\alpha < \frac{3\pi}{4}$ or
$\omega < \frac{3\pi}{4}$.}]\hspace{.2\textwidth}

Without loss of generality assume that $\alpha < \frac{3\pi}{4}$.
By induction on $(q_1,S\setminus\{q_2\})$ we get a
\mbox{$\frac{5\pi}{4}$-open} spanning path $\langle q_1,p_1,\ldots,p_k\rangle$ on
$S\setminus\{q_2\}$.
As $\angle q_2q_1p_1 \leq \alpha < \frac{3\pi}{4}$ we get a
\mbox{$\frac{5\pi}{4}$-open} spanning path $\langle q_2,q_1,p_1,\ldots,p_k\rangle$
on $S$.

\item[{\normalfont {\bf Case 2.2\ }
    Both $\alpha$ and $\omega$ are at least $\frac{3\pi}{4}$.}]\hspace{.2\textwidth}

Let $l_1$ and $l_2$ be the lines through
$q_1$ and $q_2$, respectively, and orthogonal to $\overline{q_1q_2}$.
Further let ${\cal K} = CH(S\setminus\{q_1,q_2\})$ and with $T$ we denote
the region bounded by $\overline{q_1q_2}$, $l_1$, $l_2$ and the part of $\cal K$ closer to $\overline{q_1q_2}$ (see Fig.~\ref{fig:pathCases1} (right)).

\item[{\normalfont {\bf Case 2.2.1\ }
    At least one vertex $p$ of $\cal K$ exists in $T.$ }]\hspace{.2\textwidth}

If there exist several vertices of $\cal K$ in $T$, then we choose $p$ as
the one with smallest distance to $\overline{q_1q_2}$ (see Fig.~\ref{fig:pathCases2} (left)).
Obviously the edges $\overline{q_1p}$ and $\overline{q_2p}$
intersect $\cal K$ only in $p$ and the angles $ \alpha_1$ and
$ \beta$ are each at most $\frac{\pi}{2}$ (see Fig.~\ref{fig:pathCases2} (middle)).

\item[{\normalfont {\bf Case 2.2.1.1\ }
     $ \gamma_2 > \frac{\pi}{2}$ (see Fig.\ref{fig:pathCases2} (middle)).}]\hspace{.2\textwidth}

By induction on $(p,S\setminus\{q_1,q_2\})$ we get a
\mbox{$\frac{5\pi}{4}$-open} spanning path $\langle p,p_1,\ldots,p_k\rangle$ for \linebreak
\mbox{$S\setminus\{q_1,q_2\}$}.
Moreover the smaller of $\angle q_2pp_1$ and $\angle p_1pq_1$ is at
most $\frac{2\pi-\frac{\pi}{2}}{2} = \frac{3\pi}{4}$.
Thus we get a \mbox{$\frac{5\pi}{4}$-open} spanning path
$\langle q_1,q_2,p,p_1,\ldots,p_k\rangle$ or $\langle q_2,q_1,p,p_1,\ldots,p_k\rangle$ for $S$.

\item[{\normalfont {\bf Case 2.2.1.2\ }
     $ \gamma_2 \leq \frac{\pi}{2}$ (see Fig.\ref{fig:pathCases2} (middle)).}]\hspace{.2\textwidth}

Let $y$ and $z$ be vertices of $\cal K$, with $y$ being the clock-wise
neighbor of $p$ and $z$ being the counterclockwise one
($b$ might equal $y$ and $c$ might equal $z$).
At least one of $ \alpha_1$ or $ \beta$
\mbox{is $\geq \frac{\pi}{4}$}.
Without loss of generality assume that $ \beta \geq \frac{\pi}{4}$, the other
case is symmetric.
Then $q_1,q_2,p,y$ form a convex four-gon because
 $ \alpha \geq \frac{3\pi}{4}$ and $ \beta \geq \frac{\pi}{4}$ imply that
 $\angle bpq_2$ in the four-gon $b,q_1, q_2, p$  is less than $\pi.$ Therefore also $ \gamma \leq \angle bpq_2 < \pi.$
We show that all four angles
$ \alpha_1$, $ \gamma_1$, $ \beta_2$ and $ \delta$
are at most $\frac{3\pi}{4}$.
Then we apply induction on \mbox{$(\overline{py},S\setminus\{q_1,q_2\})$}
and get a \mbox{$\frac{5\pi}{4}$-open} spanning path on $S\setminus\{q_1,q_2\}$,
which can be completed to a \mbox{$\frac{5\pi}{4}$-open} spanning path for $S,$
$\langle q_2,q_1,p,y,p_1,\ldots,p_k\rangle$ or
$\langle q_1,q_2,y,p,p_1,\ldots,p_k\rangle$, respectively.
\begin{itemize}
\item Both $ \alpha_1$ and  $ \beta_2 <  \beta$ are clearly smaller than
$\frac{\pi}{2}$, hence smaller than $\frac{3\pi}{4}$.

\item For $ \gamma_1$, the supporting line of $\overline{yp}$ must cross the segment
$\overline{q_1b}$, so that we have  $ \alpha_2 +  \gamma_1 < \pi$ (they are two angles of a
triangle). Also, $ \alpha_2 =  \alpha -  \alpha_1
\geq \frac{3\pi}{4} - \frac{\pi}{2} = \frac{\pi}{4}$, so $ \gamma_1 < \frac{3\pi}{4}$.
\item Analogously, for $\delta$, observe that the supporting line of $\overline{yp}$ must cross the segment $\overline{q_2c}$, so that we have $\omega-\beta_2 + \delta < \pi$. Also $\omega-\beta_2 \geq \frac{\pi}{4}$, so $\delta < \frac{3\pi}{4}$.
\end{itemize}

\item[{\normalfont {\bf Case 2.2.2\ }
    No vertex of $\cal K$ exists in $T$.}]\hspace{.2\textwidth}

Both, $l_1$ and $l_2$, intersect the same edge $\overline{yz}$ of
$\cal K$ (in $T$), with $y$ closer to $l_1$ than to $l_2$ (see Fig.~\ref{fig:pathCases2} (right)).
We show that the four angles $\angle yzq_1$, $\angle q_2q_1z$, $\angle yq_2q_1$ and $\angle q_2yz$ are
all smaller than $\frac{3\pi}{4}$. Then induction on $(\overline{yz},S\setminus \{q_1,q_2\})$
yields a path that can be extended to a \mbox{$\frac{5\pi}{4}$-open} path
$\langle q_2,q_1,z,y,p_1,\ldots,p_k\rangle$ or
$\langle q_1,q_2,y,z,p_1,\ldots,p_k\rangle$.
Clearly, the angles $\angle q_2q_1z$ and $\angle yq_2q_1$ are both
smaller than $\frac{\pi}{2}$.
The sum of $\angle q_2yz$ + $\angle cq_2y$ is smaller than $\pi$
because the supporting line of $\overline{yz}$ intersects the segment $\overline{q_2c}.$
Now, $\angle cq_2y$ is at least $\frac{\pi}{4}$ by the assumption that $\angle cq_2q_1 \geq \frac{3\pi}{4}.$
So, $\angle q_2yz < \frac{3\pi}{4}$.
The symmetric argument shows that $\angle yzq_1 < \frac{3\pi}{4}.$ 
\end{description}
\vspace{-1.2\baselineskip}%
\end{proof}

\noindent
It is essential for Theorem~\ref{thm:5pi4thpath} that the starting point of a $\frac{5\pi}{4}$-open path is an extreme point of $S$, as an equivalent result is in general not true for interior points. As a counter example consider a regular $n$-gon with an additional point in its center. It is easy to see that for sufficiently large $n$ starting at the central point causes a path to be at most $\pi+\epsilon$-open for a small constant $\epsilon$. Similar, non-symmetric examples exist already for $n \geq 6$ points, and analogously, if we require a specific interior edge to be part of the path, there exist examples bounding the openness by~$\frac{4\pi}{3}+\epsilon$~\cite{V}.

\section{Conclusion and open problems}

In this paper we introduced the concept of openness of plane straight line graphs, a generalization of pointedness as used in the context of pseudo-triangulations. We derived bounds for the maximal openness for the classes of triangulations, spanning trees (general, as well as with bounded vertex degree), and spanning paths. Despite the examples presented in the final discussion of Section~\ref{sec:path_general} we 
state the following conjecture:

\begin{conjecture}
  Every finite point set in general position in the plane has a
  $\frac{3\pi}{2}$-open spanning path.
\end{conjecture}
Of interest are of course also the algorithmic problems associated with openness:
for a given point set, how fast can we compute the maximal open plane straight-line 
graph of a given class? For which classes can this be done in polynomial time?

\paragraph{\bfseries Acknowledgements.}
Research on this topic was initiated at the third European
Pseudo-Tri\-an\-gu\-la\-tion working week in Berlin, organized by G\"unter
Rote and Andr\'e Schulz. We thank Sarah Kappes, Hannes Krasser, David
Orden, G\"unter Rote, Andr\'{e} Schulz, Ileana Streinu, and Louis
Theran for many valuable discussions. We also thank Sonja
\v{C}uki\'{c} and G\"unter Rote for helpful comments on the
manuscript.


\begin{thebibliography}{99}

\bibitem{AAK}
    E. Ackerman, O. Aichholzer, and B. Keszegh. 
    Improved upper bounds on the reflexivity of point sets.
    In  \emph{Computational Geometry: Theory and Applications}, \textbf{42} (2009), 241--249.

\bibitem{ACKMS} A. Aggarwal, D. Coppersmith, S. Khanna, R. Motwani,
  and B. Schieber. The Angular-Metric Traveling Salesman Problem.
  \emph{SIAM Journal on Computing} \textbf{29}, 3 (1999), 697--711.

\bibitem{AHHHSSV}
    O. Aichholzer, T. Hackl, M. Hoffmann, C. Huemer, A. Por, F. Santos, B. Speckmann, and B. Vogtenhuber. 
    Maximizing maximal angles for plane straight line graphs. 
    In  \emph{Proc. 10th  Workshop on Algorithms and Data Structures}, 
    LNCS \textbf{4619}, 458--469, 2007.

\bibitem{AABK} O. Aichholzer, F. Aurenhammer, H. Krasser, and P. Brass. 
    Pseudo-Triangulations from Surfaces and a Novel Type of Edge Flip. 
    \emph{SIAM Journal on Computing} \textbf{32}, 6 (2003), 1621--1653.

\bibitem{AFHMNSS} E. Arkin, S. Fekete, F. Hurtado, J. Mitchell, M. Noy, V. Sacrist{\'a}n, and S. Sethia. 
    On the Reflexivity of Point Sets. 
    \emph{Discrete and Computational Geometry: The Goodman-Pollack Festschrift.}
    Springer-Verlag, \emph{Series Algorithms and Combinatorics}
    \textbf{25} (2003),
    139--156.

\bibitem{AX} F. Aurenhammer and Y.-F. Xu. 
    Optimal Triangulations.
    Kluwer Academic Publishing, \emph{Encyclopedia of Optimization}, \textbf{4} (2000), 160--166.
 
\bibitem{BPV} I. B{\'a}r{\'a}ny, A. P{\'o}r, and P. Valtr. 
  Paths with no Small Angles. 
  In \emph{Proc. 8th Latin American Symposium on Theoretical Informatics}, 
  654--663, 2008.

\bibitem{BEG} M. Bern, D. Eppstein, and J. Gilbert. 
    Provably Good Mesh Generation. 
    \emph{Journal of Computer and System Sciences} \textbf{48}, 3 (1994), 384--409.

\bibitem{DKC} Y. Dai, N. Katoh, and S.-W. Cheng. 
    {LMT}-Skeleton Heuristics for Several New Classes of Optimal Triangulations.
    \emph{Computational Geometry Theory and Applications} \textbf{17}, 1--2 (2000), 51--68.

\bibitem{ETW} H. Edelsbrunner, T. S. Tan, and R. Waupotitsch. 
    An $O(n^2 \log n)$ Time Algorithm for the Minmax Angle Triangulation.
    \emph{SIAM Journal on Scientific and Statistical Computing} \textbf{13}, 4 (1992), 994--1008.

\bibitem{E} D. Eppstein. 
    The Farthest Point {D}elaunay Triangulation Minimizes Angles. 
    \emph{Computational Geometry Theory and Applications} \textbf{1}, 3 (1992), 143--148.

\bibitem{FW} S. P. Fekete and G. J. Woeginger.  
    Angle-Restricted Tours in the Plane.  
    \emph{Computational Geometry Theory and Applications} \textbf{8}, 4 (1997), 195--218.

\bibitem{HORSSSSSW} R. Haas, D. Orden, G. Rote, F. Santos, B.
    Servatius, H. Servatius, D. Souvaine, I. Streinu, and W. Whiteley.
    Planar Minimally Rigid Graphs and Pseudo-Triangulations.
    \emph{Computational Geometry Theory and Applications} \textbf{31}, 1--2 (2005), 31--61.

\bibitem{KW}  J. M. Keil and T. S. Vassilev. 
  The Relative Neighbourhood Graph is a Part of Every 30$\deg$-Triangulation.
  \emph{Information Processing Letters} \textbf{109}, 2 (2008), 93--97.

\bibitem{KSS} D. Kirkpatrick, J. Snoeyink, and B. Speckmann.  
    Kinetic Collision Detection for Simple Polygons. 
    \emph{International Journal of Computational Geometry and Applications} \textbf{12}, 1--2 (2002), 3--27.


\bibitem{RSS} G. Rote, F. Santos, and I. Streinu.
  Pseudo-Triangulations -- a Survey.
  \emph{Surveys on Discrete and Computational Geometry -- Twenty Years Later. 
    Contemporary Mathematics} \textbf{453} 
    (2008), 343--410.

\bibitem{S} I. Streinu. 
    Pseudo-Triangulations, Rigidity and Motion Planning. 
    \emph{Discrete and Computational Geometry} \textbf{34}, 4 (2005), 587--635.

\bibitem{V} B. Vogtenhuber.
    On Plane Straight Line Graphs. Master's Thesis, Graz University of Technology, Graz, Austria, January 2007.
\end{thebibliography}
\end{document}